\documentclass[journal,twoside]{IEEEtran}
\usepackage{cite}
\usepackage{amsmath,amssymb,amsfonts}
\usepackage{algorithmic}
\usepackage{graphicx}
\usepackage{textcomp}
\usepackage{xcolor}
\usepackage{verbatim} 
\usepackage{multicol}
\usepackage{multirow}

\usepackage{caption} 
\def\BibTeX{{\rm B\kern-.05em{\sc i\kern-.025em b}\kern-.08em
    T\kern-.1667em\lower.7ex\hbox{E}\kern-.125emX}}
\begin{document}

\title{Data-Driven Modulation and Antenna Classification of Wireless Digital Communication Signals\\
}

\author{\IEEEauthorblockN{Apostolos Pappas and Antonios Argyriou}\\
\IEEEauthorblockA{Department of Electrical and Computer Engineering, University of Thessaly, Volos, Greece \\
apopappas@uth.gr, anargyr@uth.gr}

}

\newcommand{\antonis}[1]{\textcolor{blue}{{AA: \rm #1}}}
\newcommand{\apostolos}[1]{\textcolor{red}{{AP: \rm #1}}}

\maketitle

\begin{abstract}
In this paper we are interested to learn from a wireless digitally modulated signal the number of antennas that the transmitter (Tx) of this signal uses, as well as its specific modulation scheme (from phase-shift keying (PSK) or quadrature amplitude modulation (QAM)). Formally, these are modulation and antenna classification problems. We examine the problems with data-driven machine learning (ML)-based techniques. The two sub-problems of modulation and number of transmitter antenna classification are initially examined independently for a variety for system parameters, namely the SNR, number of receiver (Rx) antennas, and classification algorithms. Then we consider the joint problem where we follow two approaches. One, where the sub-problems are solved independently and in parallel, and one where the antenna classifier waits on the result of the modulation classifier. The two proposed schemes do not require any knowledge/details of the used modulation schemes and the way the Tx antennas are used (spatial multiplexing, space-time codes,etc.) as it is fully data-driven and not decision-theoretic based. The results of our approach are characterized by high classification accuracy and they pave the way for more ML-based data-driven techniques that reveal more characteristics of the Tx.
\end{abstract}



\section{Introduction}
Wireless digital communication signals may be modulated with a variety of techniques. Being able to classify the modulation type used in an unknown wireless signal is usually described as Automatic Modulation Classification (AMC). AMC has several prospective applications in wireless communication. 
%
The first area where AMC has a key role is in \emph{automatic receiver configuration}. 
To demodulate the modulated signals, and to recover the transmitted message, a receiver must know the modulation type. But when the receiver is blind it has to detect the used modulation scheme by employing AMC so that is configures itself appropriately. 
%
%
The second potential application of AMC is the recovery of the modulation scheme (and possibly of the modulated data), from any arbitrary wireless digitally modulated signal. This example of applying AMC to an arbitrary transmitter (Tx), can be seen from two angles: If the Tx is an enemy (in military applications) we may use AMC to identify the type of the Tx and in this case we are interested to \emph{improve AMC performance}. On the other hand, if we are the Tx then an unauthorized Rx may employ AMC in order to identify the used modulation. In this second scenario we see that AMC is deployed to compromise the privacy of a transmitter which means that, as a transmitter, we are interested to \emph{degrade AMC performance}.




Regardless of which of the aforementioned applications we are interested, in this paper we argue that AMC is only one aspect of a bigger more interesting problem. The modulation is only one characteristic of the transmission that may be unknown. Classification of wireless communication signals is inherently a multidimensional problem. In this paper we take one more step towards the direction of learning more about a digitally modulated transmitted signal. To this aim we examine the potential of classifying in a blind scenario both the modulation type and the number of Tx antennas with data-driven machine learning (ML) algorithms.  

We investigate solving the problem with two different methods. With the first method the two sub-problems are solved independently and in parallel, i.e. modulation classification is performed without interaction with the number of antennas classifier. Hence, no intermediate result is passed on from one part of the system to the other. With the second method, the joint signal classification is performed in two sequential phases. In the first phase, the receiver is responsible for classifying the modulation of the incoming signal. Having recognized the way that the received signal is modulated, the next step is to classify the number of transmitting antennas with classifiers pre-trained and dedicated for each modulation scheme. With both schemes modulation classification is performed hierarchically. The overall system entails first baseband data pre-processing for feature extraction and dataset creation. 
Overall, the contributions of this paper are the following:
\begin{itemize}
    \item We present a new data-driven ML-based methodology for classifying independently the number of used Tx antennas and the used modulation without any type of knowledge regarding the modulations (e.g. their probability distribution function (PDF) that may be a complicated expression), and also the way Tx antennas are used (spatial multiplexing, space-time coding, selection diversity, etc.). That is our system is completely agnostic to the Tx configuration and is only empowered by the training data.
    \item We present a joint classification scheme where the two problems are treated sequentially, instead of completely independently and in parallel, leading to even higher gains in classification accuracy.  
    \item For each sub-problem on their own, but also jointly, we present a thorough evaluation of different classifiers, including new ensemble methods, over a wide range of system parameters.
\end{itemize}

The paper is organized as follows. Related work in analyzed in Section~\ref{sec:related-work} and in Section~\ref{sec:system-model} we present the communication system model and assumptions. Section~\ref{sec:classification} presents feature extraction and dataset creation methodology. Sections~\ref{sec:modulation-classification} and~\ref{sec:antenna-classification} deal with modulation and antennas classifical respectively. The joint classification problem is addressed in Section~\ref{sec:joint-classification}. Finally, \ref{sec:joint-classification} concludes this paper.

%

\section{Related Work}
\label{sec:related-work}

Due to its importance, AMC has received significant attention from the research community. AMC algorithms can be categorized into two groups, the Likelihood-Based (LB) and the Feature-Based (FB) schemes. The way in which the LB schemes address AMC is by employing methods such as the Generalized-Likelihood-Ratio-Test (GLRT), the Average-Likelihood-Ratio-Test (ALRT), the Hybrid-Likelihood-Ratio-Test (HLRT) as well as other variants of LRTs. Regarding the FB approach, a big advantage in following it is the simplicity of its implementation as well as the performance which approaches the one achieved by LB schemes. In \cite{hameed2009likelihood, dobre2007survey} the reader can find earlier detailed reports on various methods based on the LB approach. Older and more recent methods propose the use of a plethora of different features for FB algorithms, ranging from raw signal characteristics \cite{assaleh1992new, dan2005new, xie2008efficient, peng2018modulation, ahmadi2010using} to other statistically extracted features. The latter type comprises of features such as high order moments (HOMs) \cite{zhou2010signal, swami2000hierarchical}, high order cumulants (HOCs) \cite{liu2006novel}, very high order statistics (VHOS) \cite{su2013feature} and others \cite{dobre2003higher}. Consequently, the differentiation between various FB-based methods can be found mainly in the type of used features used and the way these features are extracted~\cite{o2017introduction}. Nevertheless, HOCs seem to be the most promising types of features since they differentiate very well the cyclo-stationary nature of modulated signals. If we go beyond the features themselves and focus on how they are used, a number of ML methods (both supervised and unsupervised) have been developed towards AMC \cite{meng2018automatic, wang2020lightamc, zhang2018automatic, ramjee2019fast, ali2017unsupervised, o2016convolutional}. The main results from these studies is that beyond the features themselves the classifier architecture has been shown to be important. The most promising avenue has been reported in~\cite{swami2000hierarchical,abdelmutalab2016automatic}, where hierarchical classifiers were proposed using statistical features extracted from the incoming signal. In~\cite{abdelmutalab2016automatic} in particular, a hierarchical structure is presented for classifying between six types of modulation (BPSK, QPSK, 8-PSK, 16-QAM, 64-QAM and 256-QAM). These obtained features are expanded into a higher dimensional space by a second degree polynomial fit. One important note is that although good performance is achieved in all the aforementioned classifiers, there is no study of more modern ensemble methods.

Regarding the classification on the number of antennas, fewer number of works exist. These techniques typically exploit pilot patterns in \cite{oularbi2012enumeration, oularbi2013exploiting, ohlmer2008algorithm} by considering cooperative scenarios (where the Tx/Rx exchange some information) with cognitive receivers. In a non-cooperative scheme, this type of classification was investigated with the use of the Akaike Information Criterion (AIC) and the Minimum Description Length (MDL) estimators~\cite{somekh2007detecting}. Although both AMC and antenna detection are studied separately, to the best of our knowledge, only one work examines these two problems in a joint manner. In their work \cite{turan2015joint}, Turan et al. proposed a decision theoretic approach for spatial multiplexing (SM) Multiple-Input and Multiple-Output (MIMO) systems considering the modulation classification and the antenna detection as a joint problem without Tx/Rx cooperation. The aforementioned joint problem is solved by minimizing the extended MDL criterion so as to detect the number of antennas and the used modulation. However, the system is engineered to operate for a Tx that employs SM only, while the number of Rx antennas is always larger than the number of Tx antennas. 
\section{Communication System Model} 
\label{sec:system-model}
We consider a generic model of an $N_T$$\times$$N_R$ MIMO system that can also be simplified to the MISO, SIMO, and SISO models depending on the selection of the Tx and Rx antennas, $N_T$ and $N_R$ respectively. The baseband model of the $N_R$$\times$$1$ received signal vector $\mathbf{y}$ is 
\begin{align}
\mathbf{y} = \mathbf{H}\mathbf{x} + \mathbf{n}
\label{eqn:signal-model}
\end{align}
where $\mathbf{H}$ is the $N_R$$\times$$N_T$ channel matrix. The $N_T$$\times$$1$ vector $\mathbf{x}$ contains the complex modulated symbols that are drawn from an alphabet $\mathcal{M}$. Also $\mathbf{n}$ is the vector of the narrowband Additive White Gaussian Noise (AWGN), which has power $\sigma^2$. We consider a quasi-static flat fading channel which means, first that the channel does not change within the time period that a PHY frame is transmitted, and second that the LTI response of the channel is modeled with a single tap. The Rayleigh distribution is assumed to be the statistical model of the channel response\cite{tse2005fundamentals}, i.e., each element $(i,j)$ of the channel matrix is a complex Gaussian random variable $\mathbf{H}(i,j)$$\sim$$\mathcal{CN}(0,\frac{1}{\sqrt{2}})$. For the sake of this paper we assume that the transmitted symbols in the vector $\mathbf{x}$ are all different, that is we use a Spatial Multiplexing (SM) MIMO transmission mode even though this is not a requirement for the feature creation and the ML algorithms we investigate (e.g. space-time codes can be used as well). For this SM system transmit power per symbol is normalized to unity and so the transmit signal-to-noise ratio (SNR) is defined as  $\frac{N}{\sigma^2}$~\cite{tse2005fundamentals}.
Gray encoding is used for the transmitted bits before they are converted to the baseband symbols. Both the real and imaginary parts of the baseband signal $\mathbf{x}$ are filtered through the transmitter (Tx) filter. For this work the filter is chosen to be an ideal sinc pulse allowing thus the creation of a narrowband transmitted signal of bandwidth $W$Hz. The typical next step is the conversion of the previous information into an analog waveform suitable for transmission. This is done with an digital-to-analog converter (DAC) that is followed by an analog mixer for up-conversion/modulation to the desired carrier frequency. Details of such a narrowband wireless communication system can be found in~\cite{argyriou2020}. 

For this system model captured in~\eqref{eqn:signal-model}, we created a custom simulator that was used for producing the desired datasets. Regarding the next step we can now define formally the objective of this work which is to identify $N_T$ and $\mathcal{M}$, that is the number of transmitter antennas and the modulation used by the Tx. Note that from the well-known baseband model in~\eqref{eqn:signal-model} the information we seek to identify is the only one we can obtain regarding the Tx configuration (beyond of course the baseband signal $\mathbf{x}$). In this model the receiver may employ a different number of antennas $N_R$. Hence, the problem we face is a multi-class classification problem that we investigate thoroughly with modern data-driven ML tools instead of classic decision-directed detection strategies.

\section{Feature Extraction \& Dataset Creation from Baseband Receiver Data}
\label{sec:classification}

The receiver examined in this paper does not perform any type of channel equalization or demodulation but simply down-conversion and digitization with an analog-to-digital converter (ADC). Consequently, we have a blind Rx which does note have Channel Side Information (CSI), and is also completely unaware of the Tx communication system configuration (knowledge of $N_T$ and $\mathcal{M}$). The first task performed by the receiver is feature extraction. Throughout this work, the feature extraction process is solely based on the raw I/Q data of the received signal $\mathbf{y}$. 

\subsection{Feature Extraction}
\label{feature_ex}
As we discussed in the introduction it is well known that HOCs capture very well the cyclo-stationary nature of modulated signals and the differences between different modulation types~\cite{zhou2010signal, swami2000hierarchical,liu2006novel,su2013feature,o2017introduction}. Hence, in this work HOCs were used as features for the ML and DL models. HOCs can be thought of as functions containing HOMs. In this work, HOMs extracted from the unknown received signal are calculated as:
\begin{align}
M_{pq} = \mathrm{E}[\mathbf{y}^{p-q}(\mathbf{y^*})^q]
\end{align}
The above equation describes the $p$-th order moment as computed using the complex-valued received signal. The second, fourth, and sixth order HOCs based on HOMs we used are:  
\begin{align*}
\text{SOC:}
\begin{cases}
& C_{20} = M_{20} \\
& C_{21} = M_{21}
\end{cases}
\end{align*}
\begin{equation*}
\text{FOC}:
\begin{cases}
& C_{40} = M_{40} - 3M_{20}^2 \\
& C_{41} = M_{40} - 3M_{20}M_{21}\\
& C_{42} = M_{42} + |M_{40}|^2 - 2M_{21}^2
\end{cases}
\end{equation*}
\begin{equation*}
\text{SiOC}:
\begin{cases}
& C_{60} = M_{60} - 15M_{20}M_{40} + 30M_{20}^3 \\
& C_{61} = M_{61} - 5M_{21}M_{40} -10M_{20}M_{41} + 30M^2_{20}M_{21}\\
& C_{62} = M_{62} - 6M_{20}M_{42} - 8M_{21}M_{41} - M_{22}M_{40} \\
&+ 6M_{20}^2M_{22} + 24M_{21}^2M_{20}\\ 
& C_{63} = M_{63} - 9M_{21}M_{42} + 12M_{21}^3 - 3M_{20}M_{43}\\
& - 3M_{22}M_{41} + 18M_{20}M_{21}M_{22}  
\end{cases}
\end{equation*}
As proposed in \cite{geisinger2010classification}, HOCs are normalized having each cumulant raised to the power of $\frac{2}{p}$. Additionally, the magnitudes of the corresponding cumulants are used as actual input to the ML and DL models~\cite{abdelmutalab2016automatic}.

\subsection{Dataset Creation}
\label{dataset_creation}
A dataset was produced for different combinations of $N_T,N_R,\mathcal{M}$. $N_T$ varied from 1, 2, and 4 Tx antennas, and six different modulations were considered (BPSK, QPSK, 8PSK, 16QAM, 64QAM, 256QAM). Each dataset was produced by transmitting 1024 symbols with the samples per symbol set to one. A number of 600 samples out of each modulation and Tx antennas dataset we taken for training. In addition, the number of channel realizations is the same for each group of transmitted symbols. We also created datasets for $N_R$=1, $N_R$=2. The produced dataset have the format illustrated in Table \ref{table:dataset_description} and their total number is 3$\times$6$\times$2=36.\footnote{We plan to share these datasets with the publication of this article.}

\begin{table*}[h!]
\centering
 \begin{tabular}{|l|l|l|l|l|l|l|l|l|l|l|l|} 
 \hline
 $C_{20}$ & $C_{21}$ & $C_{40}$ & $C_{41}$ & $C_{42}$ & $C_{60}$ & $C_{61}$ & $C_{62}$ & $C_{63}$ & Modulation & Tx antennas \\ [0.5ex] 
 \hline
   \multicolumn{9}{|c|}{Numerical (Used as input features)}&\multicolumn{2}{c|}{Nominal (Used as Classes)}\\
 \hline
 \end{tabular}
 \caption{\small Dataset Format.}
\label{table:dataset_description}
\end{table*}
\section{Modulation Classification with The Hierarchical Scheme}
\label{sec:modulation-classification}

\subsection{The Hierarchical Scheme with Different Branch Classifiers}
Due to the hierarchical structure, the multi-class classification procedure is split into several binary detection problems. First and foremost, a division is made into the two main modulation classes, namely PSK and QAM. This is the first step of the classification.
In the second level, the PSK branch classifier is responsible of distinguishing between BPSK and the rest of the PSK modulations treated as a single class. Respectively, in the QAM branch the corresponding classifier distinguishes between 16-QAM and the rest of QAM modulation types in the same manner. In case the signal is classified as BPSK or 16-QAM, the classification procedure is completed. Otherwise, an additional binary classification is performed between QPSK and 8-PSK for the PSK branch or 64-QAM and 256-QAM for the QAM one.    

Different branch classifiers are considered. First, the polynomial classifier proposed in \cite{abdelmutalab2016automatic} is examined. 
Additionally, tests were performed using the k-Nearest-Neighbors (kNN) algorithm~\cite{cover1967nearest}. kNN is a non-parametric method for classification and regression. With kNN an object is classified by a majority vote of its \textit{k} neighbors and it is assigned to the class that is most common amongst them. 

Moreover, a number of ensemble ML algorithms were investigated as well. Ensemble methods use multiple but finite individually trained classifiers to obtain better predictive performance. 
The Random Forest (RF) algorithm is an ensemble method suitable for tasks such as classifying objects 
that consists of multiple decisions trees. Using bagging and feature randomness when constructing each tree, the algorithm aims to create an uncorrelated forest of trees whose prediction by committee is more accurate than that of any individual tree. Extra-Trees (ET), proposed in \cite{geurts2006extremely}, is another ensemble ML algorithm that was used. Similarly to the RF algorithm, it also combines the predictions from numerous decision trees, with the difference between them found in the sampling approach and the selection of cut points in order to split nodes. The third ensemble method used in our experiments was Adaptive-Boosting (AdaBoost). AdaBoost can be used concurrently with other classification algorithms, also mentioned as \textit{weak} learners. These weak models are added sequentially and trained using the weighted training data until a given number of learners are created. In the testing phase, predictions are made by calculating the weighted average of the weak models. 


\subsection{Performance Evaluation}
It is important to clarify that even though the number of Tx antennas is present in each dataset, for the modulation classification problem the class is set as the \textit{Modulation} column, that is the \textit{Tx Antennas} column in Table \ref{table:dataset_description} is disregarded in this case since it is a problem that will be considered in the next section. Regarding the configuration of the considered branch classifiers, kNN was used with the number of neighbors set equal to one. The RF and ET models were build having each one of them 100 estimators, while the AdaBoost used Decision Trees (having a max depth of one) as their base method. Each of them had a maximum number of estimators equal to 100.

\begin{figure}[h!]
\centering
 \includegraphics[width = \linewidth,]{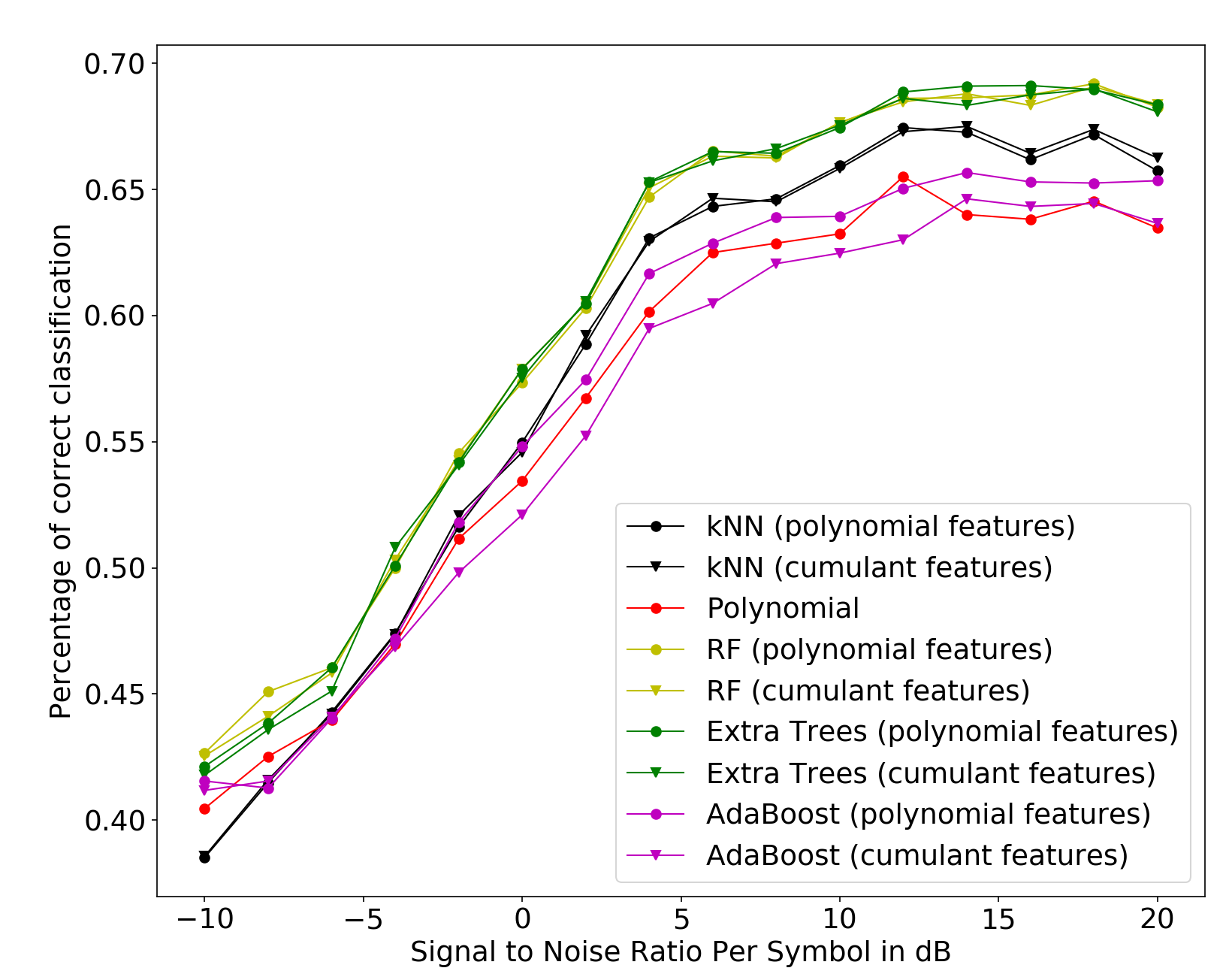}
\caption{ Percentage of correct classification plot for $N_R$=1.}
\label{fig:plot_1ant}
\end{figure}

\begin{figure}[h!]
\centering
 \includegraphics[width = \linewidth,]{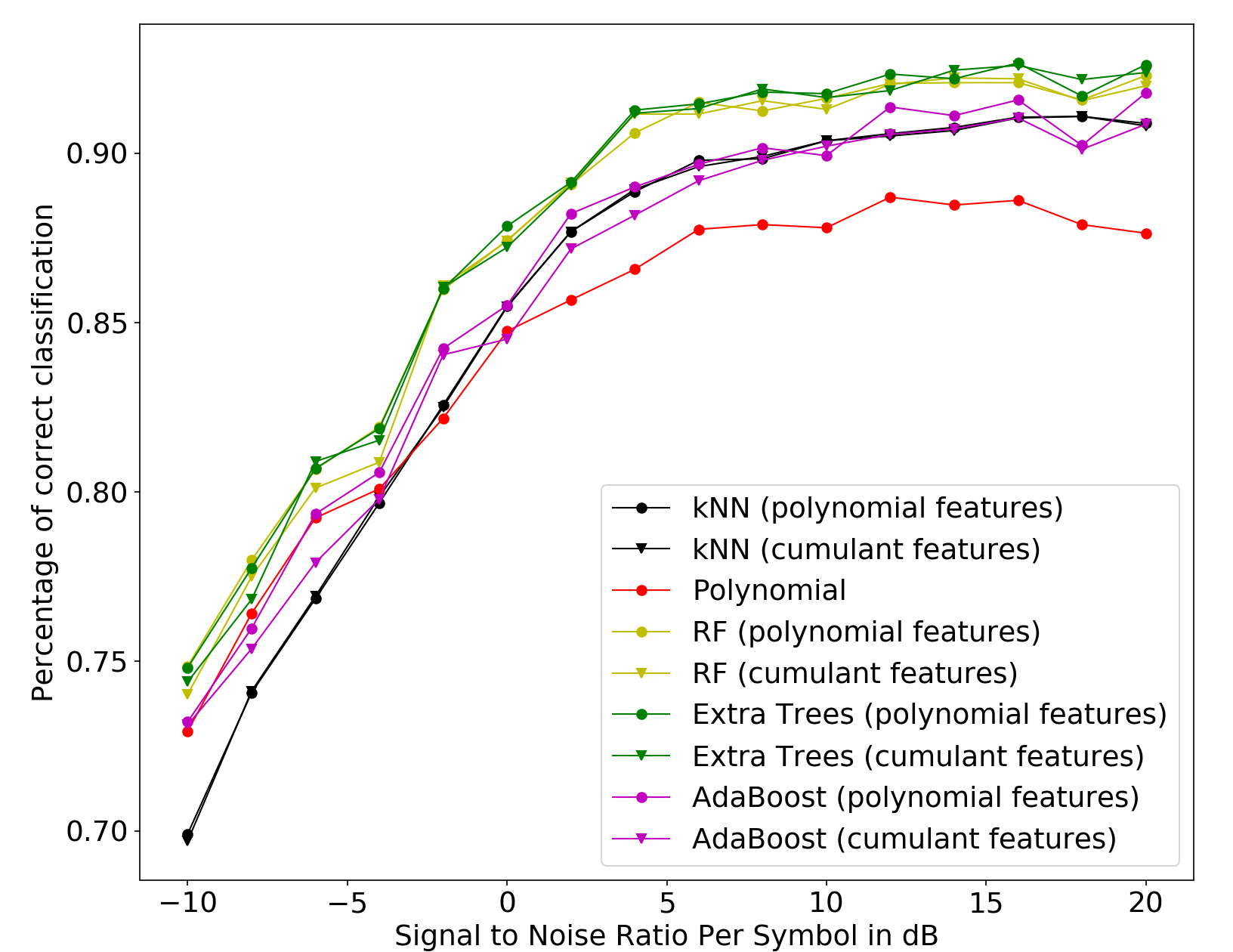}
\caption{ Percentage of correct classification plot for $N_R$=2.}
\label{fig:plot_2ant}
\end{figure}

After splitting the dataset into 60\% for training and 40\%  for testing, we fitted and tested the suggested algorithms using as input either the extracted cumulants or the extended feature vectors (referred to as polynomial features). Their performance was evaluated in terms of accuracy. Indicative accuracy scores for each algorithm on $N_R$=1 and $N_R$=2 can be found in Tables \ref{table:acc_1ant} and \ref{table:acc_2ant} respectively. It is evident that all classifiers performed much better for $N_R$=2, where the receiver is equipped with two antennas. The reason for this behavior is the symbol interference that the receiver suffers in for $N_R$=1. When the receiver has only one antenna available, symbol interference is inevitable in the cases where the transmitter uses 2 and 4 antennas. For example in the case where 2 antennas are present in the transmitter we have a 2$\times$1 MISO channel, and the received signal is written as the superposition of two transmitted symbols: $y = h_{1}x[1] + h_{2}x[2]$. Respectively, when $N_T$=4 the received signal $y$ is the sum of four interfering symbols. This means that 66\% of th dataset with $N_R$=1 contains features extracted from incoming interfered symbols, something that alters dramatically the statistical properties of each symbol. On the other hand, the dataset with $N_R$=2 possesses a much lower percentage of interfering symbols at 33\%. Regarding the accuracy of each classifier the results can be seen in Fig.~\ref{fig:plot_1ant} and Fig.~\ref{fig:plot_2ant}. The classifiers have similar performance in for $N_R$=1, with the worst one being AdaBoost using raw cumulant features. RF and ET algorithms achieve an almost equal performance either using raw cumulants or the polynomial features. For $N_R$=2, the classifiers achieve higher accuracy scores, while maintaining the same behavior. The Polynomial classifier proposed in \cite{abdelmutalab2016automatic} reports a lower saturation point. At the same time, RF and ET classifiers achieve scores up to 92\%.    


\begin{table}[t]
\centering
 \begin{tabular}{|l|l|l|l|l|} 
 \hline
 Classifier & -10dB & 0dB & 10dB & 20dB \\ [0.5ex] 
 \hline\hline
 Polynomial & 40.46  & 53.45  & 63.24  & 63.47  \\ 
 kNN (Polynomial Features) & 38.5  & 54.97  & 65.74  &  65.74  \\
 kNN (Cumulant Features) & 38.56  & 54.58  & 65.83  & 66.25   \\
 RF (Polynomial Features) & 42.66  & 57.89  & 67.66  & 68.26  \\
 RF (Cumulant Features) & 42.54  & 57.89  & 67.66  & 68.37  \\ 
 ET (Polynomial Features) & 42.12  & 57.89  & 67.45  & 68.35  \\
 ET (Cumulant Features) & 41.78  & 57.52  & 67.54  & 68.07  \\
 AdaBoost (Polynomial Features) & 41.55  & 54.81  & 63.93  & 65.34    \\
 AdaBoost (Cumulant Features) & 41.18  & 52.1  & 62.47  & 53.65\\[1ex] 
 \hline
 \end{tabular}
 \caption{Modulation classification accuracy for $N_R$=1.}
\label{table:acc_1ant}
\end{table}

\vspace{1cm}

\begin{table}[t]
\centering
 \begin{tabular}{|l|l|l|l|l|} 
 \hline
 Classifier & -10dB & 0dB & 10dB & 20dB \\ [0.5ex] 
 \hline\hline
 Polynomial & 72.93  & 84.74  & 87.8  & 87.83  \\ 
 kNN (Polynomial Features) & 69.9  & 85.48  & 90.37  &  90.87  \\
 kNN (Cumulant Features) & 69.69  & 85.46  & 90.37  & 90.81   \\
 RF (Polynomial Features) & 74.86  & 87.43  & 91.62  & 92.29  \\
 RF (Cumulant Features) & 74.02  & 87.4  & 91.29  & 91.99  \\ 
 ET (Polynomial Features) & 74.81  & 87.84  & 91.75  & 92.31  \\
 ET (Cumulant Features) & 74.39  & 87.22  & 91.64  & 92.38  \\
 AdaBoost (Polynomial Features) & 73.21  & 85.5  & 89.9  & 91.78    \\
 AdaBoost (Cumulant Features) & 73.12  & 84.51  & 90.2  & 90.85\\[1ex] 
 \hline
 \end{tabular}
 \caption{ Modulation classification accuracy for $N_R$=2.}
\label{table:acc_2ant}
\end{table}

\textit{Overall, we notice in this section that the hierarchical classification scheme can achieve even better performance with different and more sophisticated branch classifiers, epsecially the ensemble methods. Furthermore, $N_R$ increase leads to significant improvement in the the AMC performance, independently of the corresponding classification algorithm.}
\section{Antenna Classification with and without knowledge of the Modulation}
\label{sec:antenna-classification}
In this section we study Tx antenna classification. Through this scenario, the blind receiver detects the number of Tx antennas using only the raw I/Q data of the received signal. Two approaches are considered. In the first, the modulation type is totally ignored and focus is given to the classification of the number of the antennas only. Thus, we name this classifier as the Universal Antenna Classifier (UAC). In the second, the modulation type is considered known, and so the number of Tx antennas classification is performed in a stream of data modulated with the same scheme. In the second scheme, the classifiers are dedicated to each one of the investigated modulations. 




\subsection{Performance Evaluation}

Using the datasets mentioned in \ref{dataset_creation}, the goal of this section is to examine the ability of each proposed classification scheme in detecting  $N_T$. For both schemes, the kNN classifier was used, and was configured with the number of neighbors equal to 5. RF was tested, having 100 estimators, while ET was equipped with 200. Finally, AdaBoost was used having RF with 100 estimators as its base method. 

\subsubsection{The Universal Classification Scheme}
\label{Universal}
\begin{figure}[t]
	\centering
	\includegraphics[width = \linewidth,]{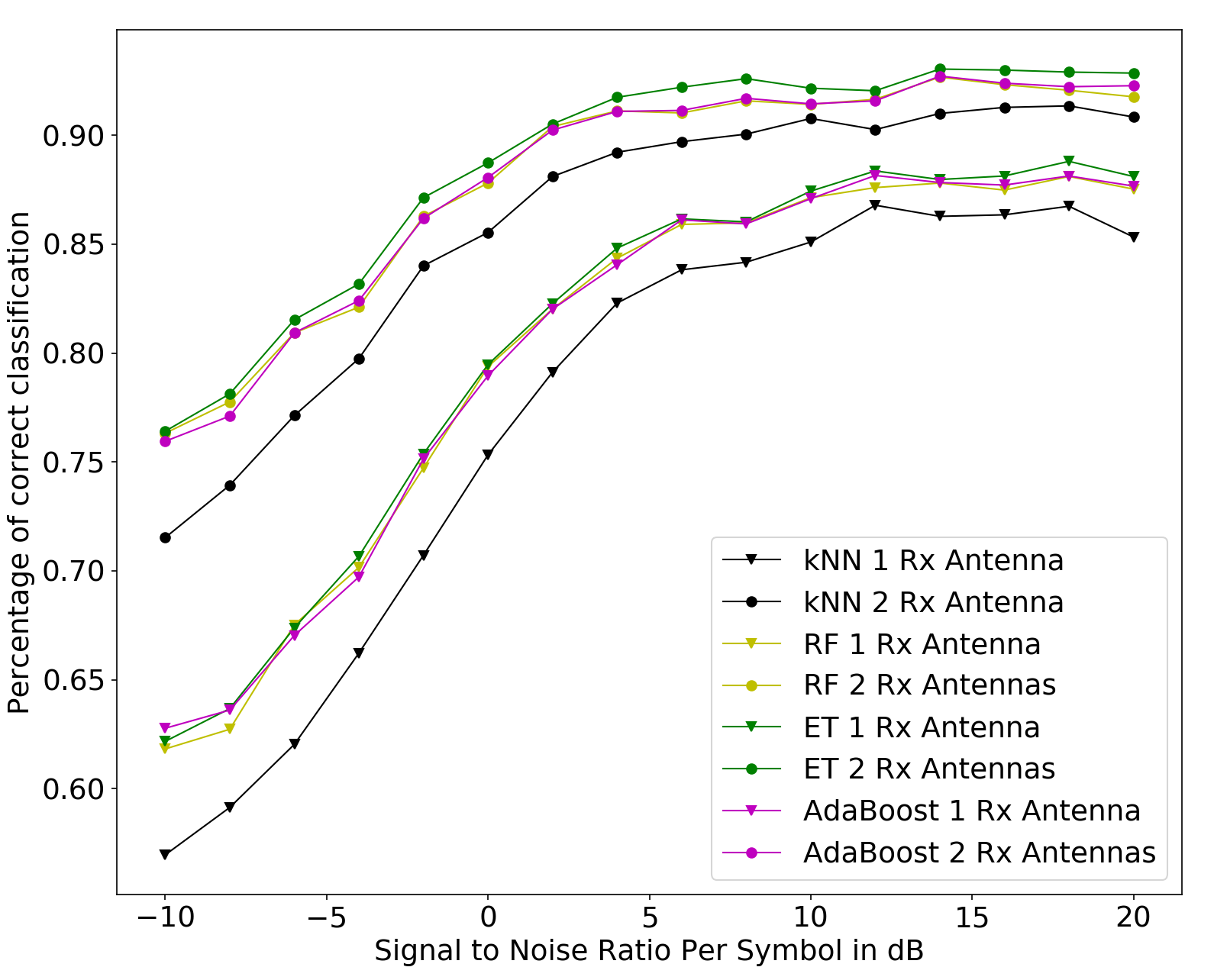}
	\caption{Percentage of correct classification with the UAC.}
	\label{fig:plot_universal_acc}
\end{figure}

Using a train-test split of 60\%-40\%, the classification algorithms were trained on 6480 samples and tested on 4320. It is evident that the ensemble methods (RF, ET, AdaBoost) performed considerably better than the kNN classifier in both datasets especially in the low SNR regime. Examining the percentage of correct classification in Fig. \ref{fig:plot_universal_acc}, we re-validate the result that multiple Rx antennas contribute to a higher accuracy. This is more evident in low SNR values, such as -10dB, where we observed an increase in accuracy up to 14.56\%. In the higher SNR regime, such as 20dB, there is still an important increase of 5.51\% (in the kNN classifier). In  table~\ref{table:classification-uac}, the reader can have an analytical report of the accuracy scores in a variety of SNR values.

\begin{table}[t]
	\centering
	\begin{tabular}{|l|l|l|l|l|l|} 
		\hline
		Dataset & Classifier & -10dB & 0dB & 10dB & 20dB \\ [0.5ex] 
		\hline\hline
		\multirow{4}{*}{$N_R$=1} 
		&kNN  & 56.96  & 75.32  & 85.09  & 85.32   \\
		&RF  & 61.82  & 79.37  & 87.12  & 87.52  \\ 
		&ET  & 62.17  & 79.46  & 87.43  & 88.1  \\
		&AdaBoost  & 62.77  & 78.95  & 87.08  & 87.66\\[1ex] 
		\hline
		\hline
		\multirow{4}{*}{$N_R$=2} 
		&kNN  & 71.52  & 85.53  & 90.76  & 90.83   \\
		&RF  & 76.31  & 87.8  & 91.41  & 91.75  \\ 
		&ET  & 76.41  & 88.72  & 92.59  & 92.84  \\
		&AdaBoost  & 75.94  & 88.05  & 91.43  & 92.26\\[1ex] 
		\hline
	\end{tabular}
	\caption{ Classification accuracy for the UAC.}
	\label{table:classification-uac}
\end{table}

\subsubsection{Dedicated Tx Antennas Classifiers}
\begin{figure}[h!]
\centering
 \includegraphics[width = 8cm, height = 8cm ]{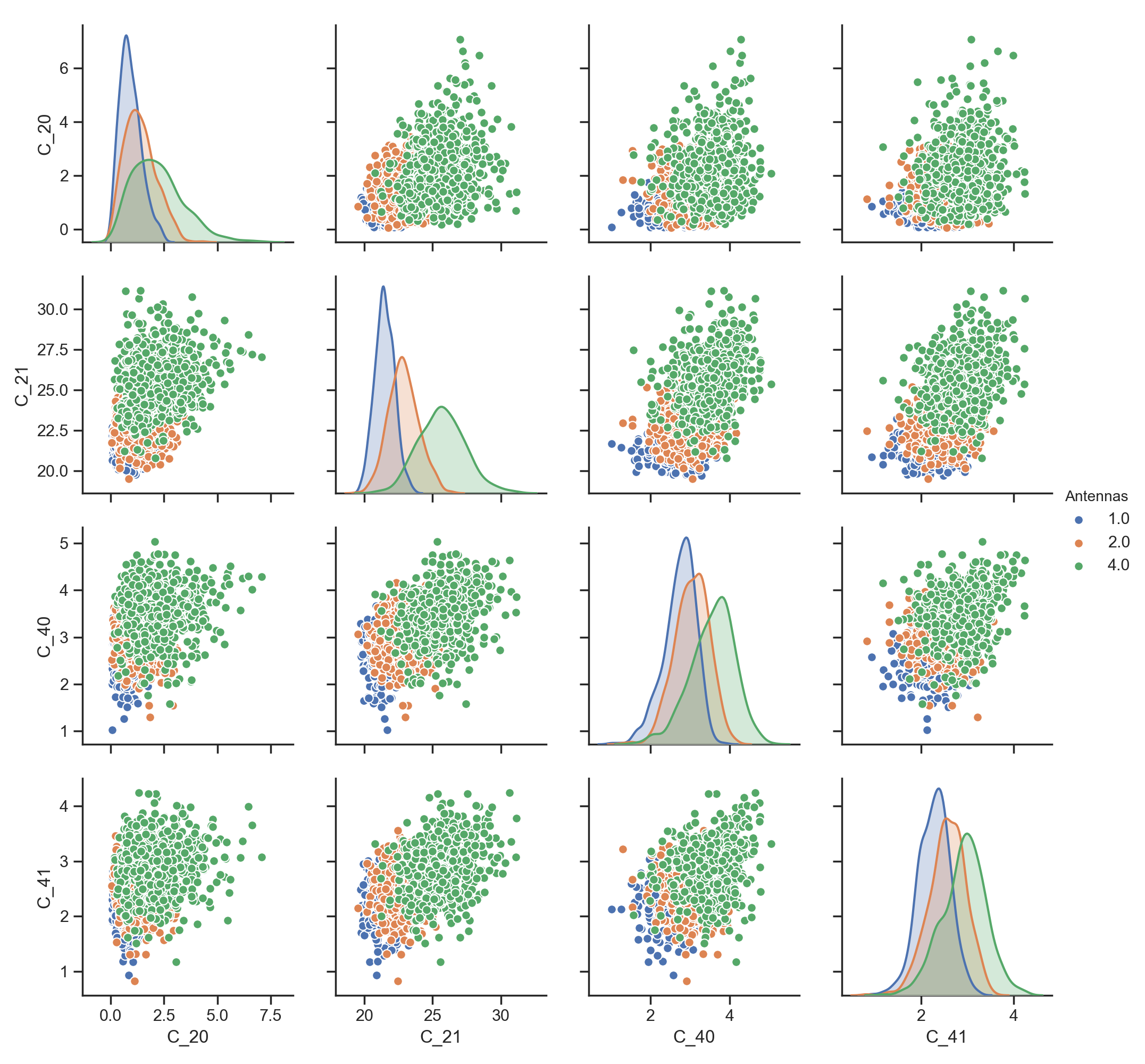}
\caption{ Pair plot of features found in BPSK modulated signals}
\label{fig:pairplotBPSK}
\end{figure}

\begin{figure}[h!]
\centering
 \includegraphics[width = 8cm, height = 8cm]{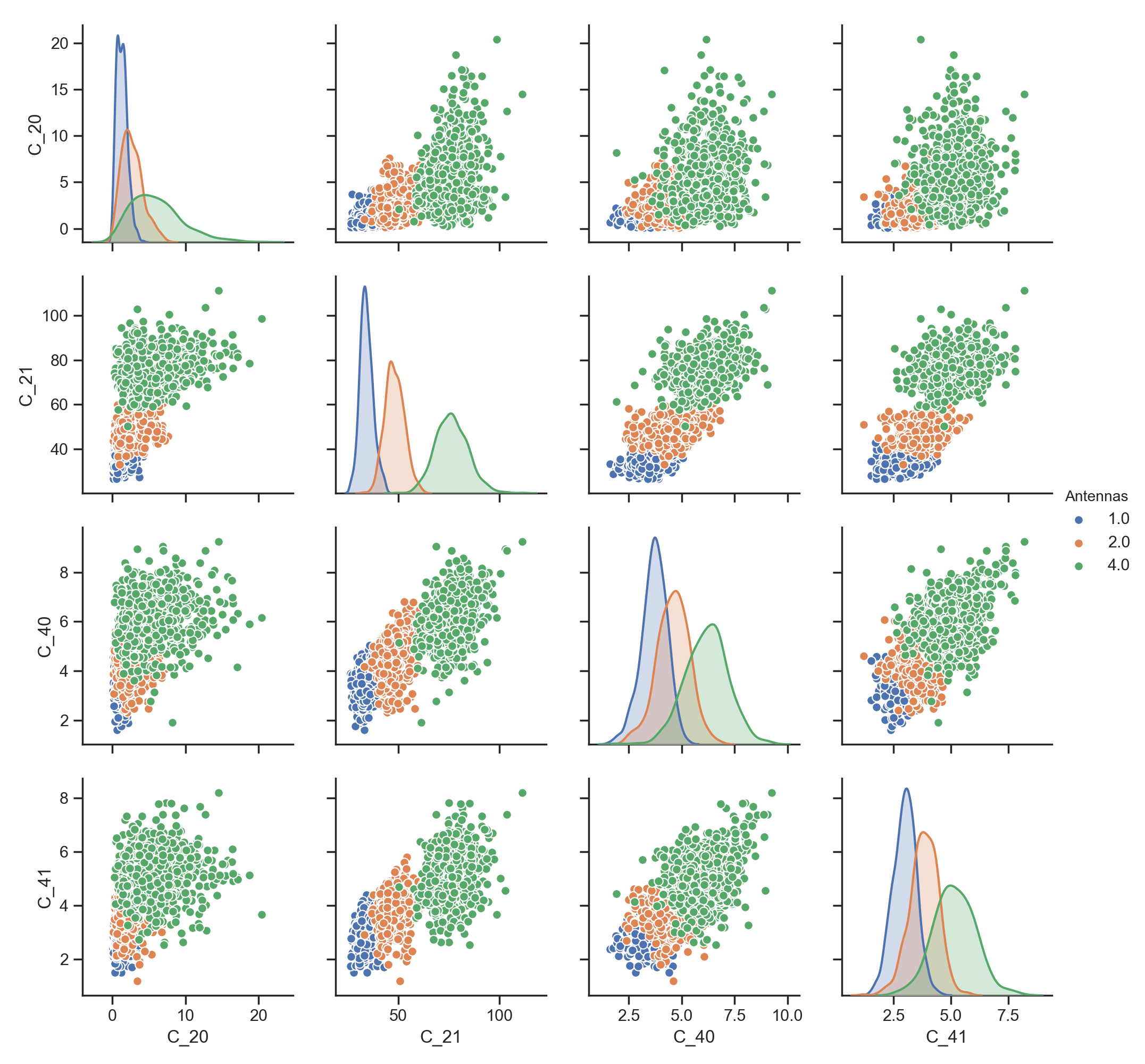}
\caption{ Pair plot of features found in 16QAM modulated signals.}
\label{fig:pairplot16QAM}
\end{figure}

Even though the Universal classifier proposed in the previous subsection performed considerably well, multiple classifiers dedicated to each of the available modulation schemes were tested. Using the same split between train and test set as in the Universal scheme, fewer data are used for training since each classifier is trained on 1080 samples and tested on 720. The classifiers were evaluated both for $N_R$=1 and $N_R$=2 and analytical accuracy reports can be found in Tables \ref{table:acc_1ant_PSK} to \ref{table:acc_2ant_QAM}. The use of two antennas by the receiver benefits the QAM modulations to a great extent (Table \ref{table:acc_2ant_QAM}) especially for lower SNR values where a rise of up to 45.14\% can be observed for -10dB. What can also be seen is that 16QAM modulation achieves a very high accuracy even with the presence of only one antenna in the receiver as seen in Table~\ref{table:acc_1ant_QAM}. Pair-plotting the corresponding features when the SNR is equal to $-10$dB (Fig. \ref{fig:pairplot16QAM}), proves that the features are in fact more distinguishable in the case of 16QAM (as well as their distributions over the available classes) than the ones derived from a BPSK modulated signal (Fig. \ref{fig:pairplotBPSK}). Comparing these two pair plots, it is easy to see that, through their distributions, the 16QAM features are more likely to be classified correctly.

\begin{table}[t]
\centering
 \begin{tabular}{|l|l|l|l|l|l|} 
 \hline
Modulation & Classifier & -10dB & 0dB & 10dB & 20dB \\ [0.5ex] 
 \hline 
\multirow{4}{*}{BPSK} 
& KNN & 55.5 & 90.5 & 97.08 & 97.08\\
& RF & 60.55 & 93.47 & 97.63 & 97.22\\
& ET & 60.55 & 93.19 & 97.63 & 98.05\\
& AdaBoost & 60.55 & 92.91 & 97.36 & 97.5\\
 \hline
 \multirow{4}{*}{QPSK} 
& KNN & 61.94 & 95 & 97.91 & 97.22\\
& RF & 64.44 & 95.69 & 97.91 & 97.36\\
& ET & 63.33 & 95.83 & 98.19 & 97.63\\
& AdaBoost & 63.19 & 95.83 & 97.5 & 97.91\\
\hline
\multirow{4}{*}{8-PSK} 
& KNN & 72.08 & 93.75 & 95.83 & 95.13\\
& RF & 78.47 & 94.44 & 95.69 & 96.11\\
& ET & 78.05 & 94.72 & 96.25 & 95.83\\
& AdaBoost & 77.63 & 94.44 & 96.66 & 96.25\\
\hline
 \end{tabular}
 \caption{Classification accuracy for $N_R$=1 and for PSK dedicated classifiers.}
\label{table:acc_1ant_PSK}
\end{table}

\begin{table}[h!]
\centering
 \begin{tabular}{|l|l|l|l|l|l|} 
 \hline
Modulation & Classifier & -10dB & 0dB & 10dB & 20dB \\ [0.5ex] 
 \hline 
\multirow{4}{*}{16-QAM} 
& KNN & 86.52 & 94.72 & 96.52 & 96.94\\
& RF & 91.66 & 94.86 & 96.94 & 97.91\\
& ET & 91.38 & 95.69 & 97.36 & 97.77\\
& AdaBoost & 91.11 & 95.13 & 97.08 & 97.91\\
 \hline
 \multirow{4}{*}{64-QAM} 
& KNN & 61.94 & 95 & 97.91 & 97.22\\
& RF & 64.44 & 95.69 & 97.91 & 97.36\\
& ET & 63.33 & 95.83 & 98.19 & 97.63\\
& AdaBoost & 63.19 & 95.83 & 97.5 & 97.91\\
\hline
\multirow{4}{*}{256-QAM} 
& KNN & 51.8 & 90.13 & 96.25 & 96.52\\
& RF  & 56.52 & 90.97 & 95.97 & 97.22\\
& ET & 57.63 & 92.22 & 95.55 & 96.94\\
& AdaBoost & 57.63 & 91.52 & 95.83 & 97.08\\
\hline
 \end{tabular}
 \caption{Classification accuracy for $N_R$=1 and for QAM dedicated classifiers.}
\label{table:acc_1ant_QAM}
\end{table}

\begin{table}[h!]
\centering
 \begin{tabular}{|l|l|l|l|l|l|} 
 \hline
Modulation & Classifier & -10dB & 0dB & 10dB & 20dB \\ [0.5ex] 
 \hline 
\multirow{4}{*}{BPSK} 
& KNN & 52.77 & 94.16 & 98.47 & 99.3\\
& RF & 60.41 & 94.58 & 98.88 & 99.58\\
& ET & 60.69 & 95 & 98.88 & 99.58\\
& AdaBoost & 60.55 & 96.66 & 98.75 & 99.58\\
 \hline
 \multirow{4}{*}{QPSK} 
& KNN & 65.55 & 97.83 & 99.58 & 99.3\\
& RF & 72.63 & 97.91 & 99.16 & 99.3\\
& ET & 71.25 & 98.05 & 99.58 & 99.44\\
& AdaBoost & 71.25 & 97.77 & 99.3 & 99.16\\
\hline
\multirow{4}{*}{8-PSK} 
& KNN & 73.33 & 97.63 & 98.47 & 98.33\\
& RF & 79.72 & 97.63 & 98.75 & 98.88\\
& ET & 81.66 & 98.19 & 98.61 & 99.44\\
& AdaBoost & 80.55 & 97.77 & 98.61 & 98.88\\
\hline
 \end{tabular}
 \caption{ Classification accuracy for $N_R$=2 and for PSK dedicated classifiers.}
\label{table:acc_2ant_PSK}
\end{table}

\begin{table}[h!]
\centering
 \begin{tabular}{|l|l|l|l|l|l|} 
 \hline
Modulation & Classifier & -10dB & 0dB & 10dB & 20dB \\ [0.5ex] 
 \hline 
\multirow{4}{*}{16-QAM} 
& KNN & 92.22 & 97.77 & 98.61 & 98.33\\
& RF & 94.44 & 98.19 & 98.75 & 98.75\\
& ET & 94.44 & 98.61 & 98.88 & 98.61\\
& AdaBoost & 94.3 & 98.33 & 98.61 & 98.61\\
 \hline
 \multirow{4}{*}{64-QAM} 
& KNN & 96.52 & 96.8 & 98.19 & 96.94\\
& RF & 98.05 & 98.05 & 98.88 & 98.05\\
& ET & 98.33 & 98.33 & 98.75 & 98.19\\
& AdaBoost & 97.77 & 97.91 & 98.8 & 98.05\\
\hline
\multirow{4}{*}{256-QAM} 
& KNN & 96.94 & 96.38 & 97.91 & 97.91\\
& RF  & 98.47 & 98.19 & 99.16 & 98.61\\
& ET & 98.33 & 98.47 & 99.44 & 98.88\\
& AdaBoost & 98.33 & 98.05 & 99.16 & 98.61\\
\hline
 \end{tabular}
 \caption{ Classification accuracy for $N_R$=2 and for QAM dedicated classifiers.}
\label{table:acc_2ant_QAM}
\end{table}

%
%

\textit{It is evident, throughout the course of this section, that antenna classification is affected by different aspects of the system. Different classification algorithms, as well as a diversity of the number of Rx antennas, improve the classification performance substantially. Furthermore, our results show that ensemble algorithms offer a better accuracy over the remaining algorithms. We also showed again that even with one Rx antenna we are able to classify with sufficient accuracy the value of $N_T$.}
\section{Joint Classification}
\label{sec:joint-classification}
In this section, the joint antenna and modulation classification problem is examined.

\subsection{Joint Classification Using the Universal Classifier}
\label{universal}
\begin{figure}[t]
\centering
 \includegraphics[width = 8cm,height = 10cm]{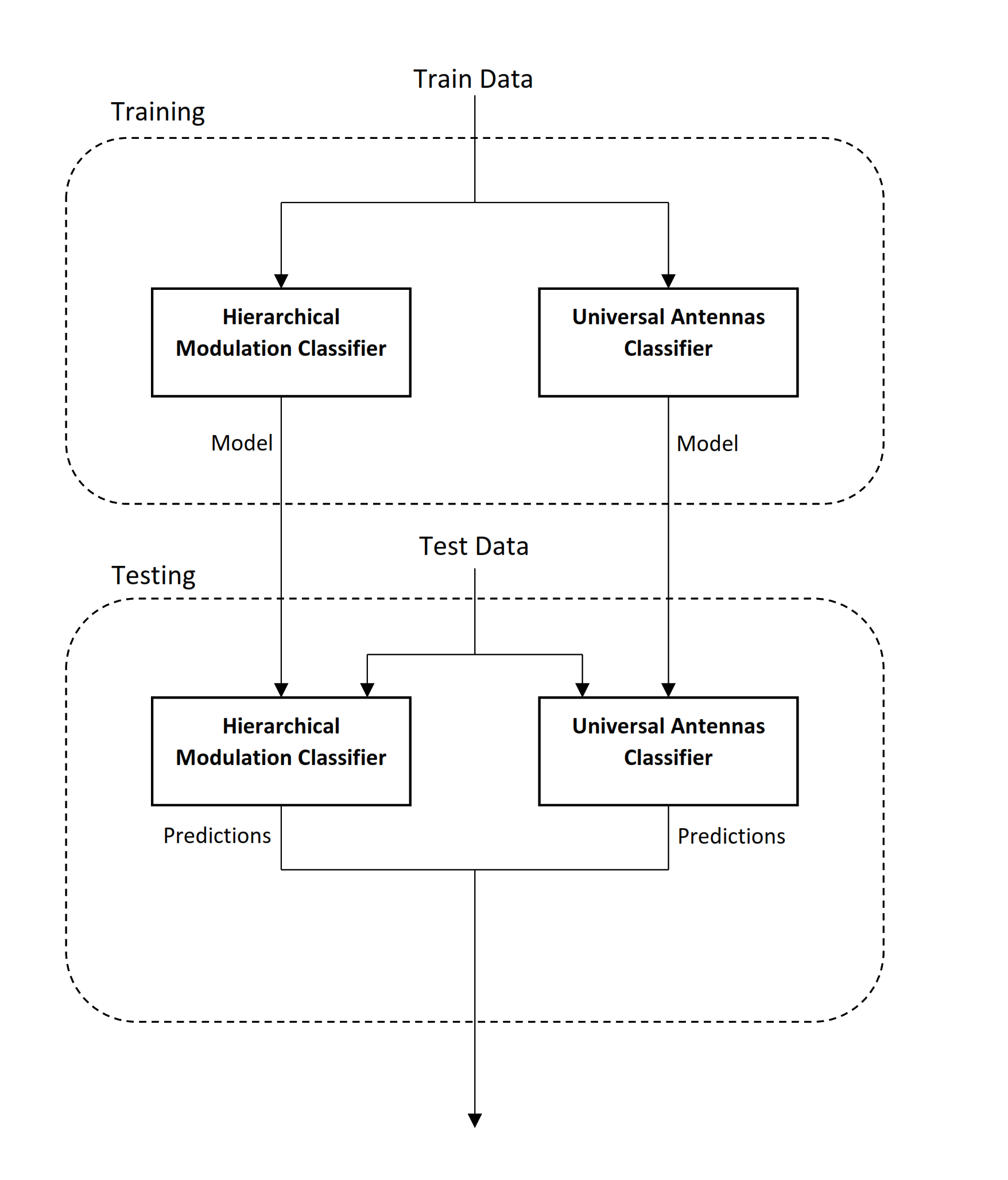}
\caption{The proposed joint classifier when using the Universal Antenna Classifier.}
\label{fig:joint_universal}
\end{figure}

In this type of joint classification the two sub-problems are treated in parallel. This is because antenna classification is not affected by the modulation classification results and vice versa. The block diagram of the proposed joint universal classifier can be seen in Fig. \ref{fig:joint_universal}. During the training phase, both classifiers are provided with the same training data. What differentiates their respective training processes, are the corresponding target values. In the case of the Hierarchical Modulation Classifier, the target values in the training procedure are present in the \textit{Modulation} column of each used dataset. During this procedure, the \textit{Antennas} column is ignored. At the same time the training of the antennas classifier is taking place, this time using the values of the \textit{Antennas} column as target values.
%
Predictions are made by each model separately. The Hierarchical Modulation model will predict the modulation type of each received signal. At the same time, the Universal Antenna Classifier model makes predictions on the number of Tx antennas for each signal contained in the test set. Then, the predictions are combined in a two column matrix and compared with the true corresponding values. In this work, we used the same type of classification algorithm for both the modulation and antennas classifiers.

\subsection{Joint Classification Using Dedicated Classifiers}
As discussed in Section~\ref{sec:antenna-classification}, the antenna classification problem can be addressed by creating classifiers dedicated to each modulation type present. Unlike the Universal approach of the previous section, where one model is built for classifying the number of antennas regardless of the way the signal is modulated, in this approach six different models for antenna classification are being created. The training stage is independent for the two different classifier types ($N_T$ and $\mathcal{M}$). As in Section \ref{universal}, the Hierarchical Modulation Classifier is trained, having the \textit{Modulation} column as its target values. On the other hand, the training of each dedicated classifier requires the fitting of six different models (that is for each type of modulation). In this way, six dedicated antenna classifiers are constructed.

\begin{figure}[t]
\centering
 \includegraphics[width = 8cm, height = 10cm]{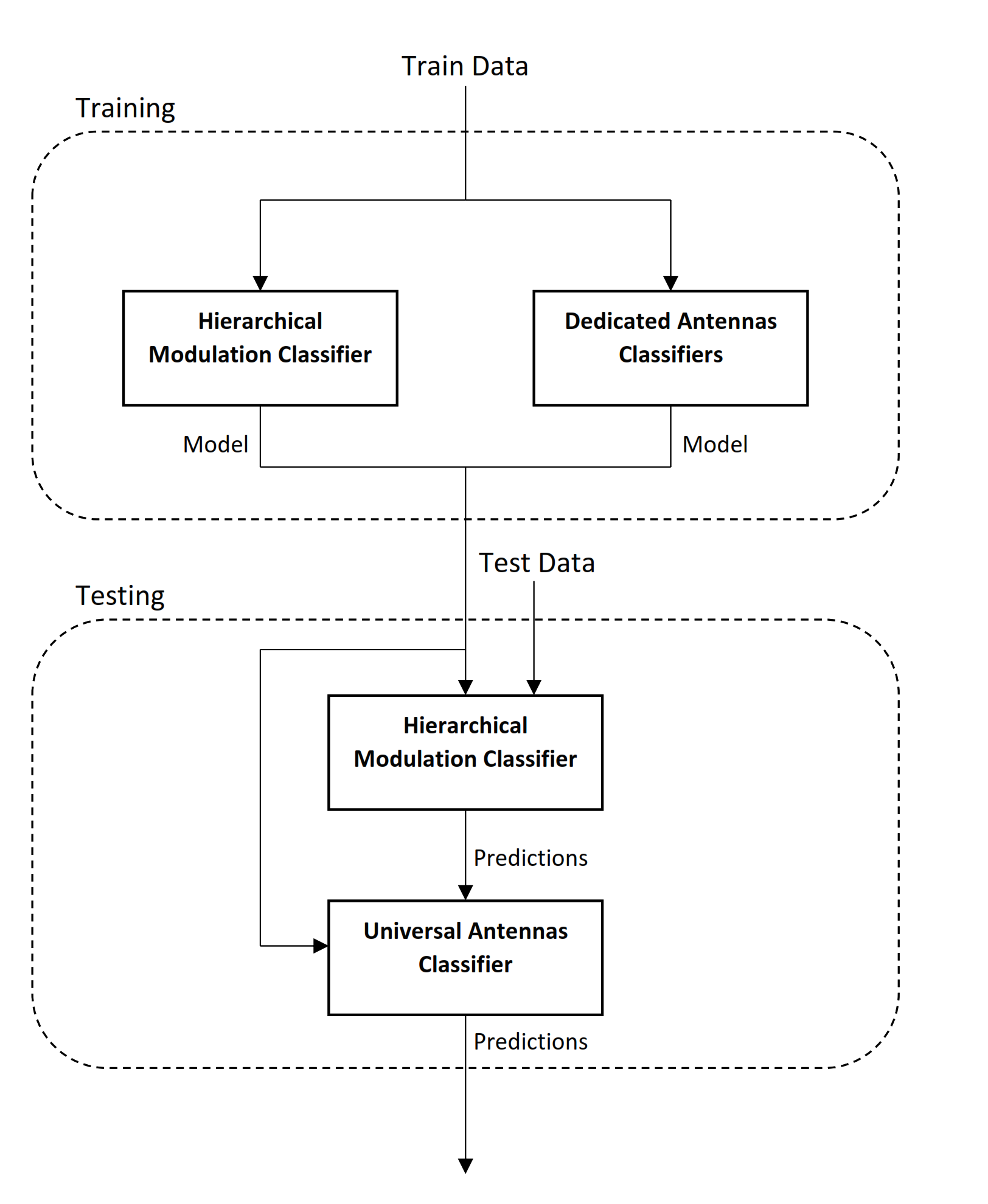}
\caption{The proposed joint classifier using the Dedicated Antenna Classifier.}
\label{fig:joint_dedicated}
\end{figure}

Testing is illustrated in the block diagram of Fig.~\ref{fig:joint_dedicated}. First, modulation classification takes place, and after the predictions made by the corresponding model the test set is split into six parts depending on the decided modulation type. It should be noted that for the dedicated models of the $N_T$ classification to work properly, data with constant modulation type should be used as input. It is then up to the selected antenna classification model to effectively classify the number of antennas $N_T$. Finally, the predictions for the number of antennas are combined with the modulation classification predictions to form a two-column matrix of the results.

\subsection{Performance Evaluation}
As in the previous section, $N_R$=1 and $N_R$=2 are used in order to provide insight into whether a diversity in the number of Tx antennas leads to better classification accuracy. The train-test split is set to 60\%-40\% as in every experiment run in the course of this paper. The classification algorithms are chosen to be kNN with 100 neighbours, RF with 100 estimators, ET with 200 estimators and AdaBoost classifier having 100 estimators and the previously described RF model as its base estimator. These models are used both in the part of Hierarchical Modulation and the Antennas classifiers, universal or dedicated.


\begin{figure}[t]
\centering
 \includegraphics[width = \linewidth,]{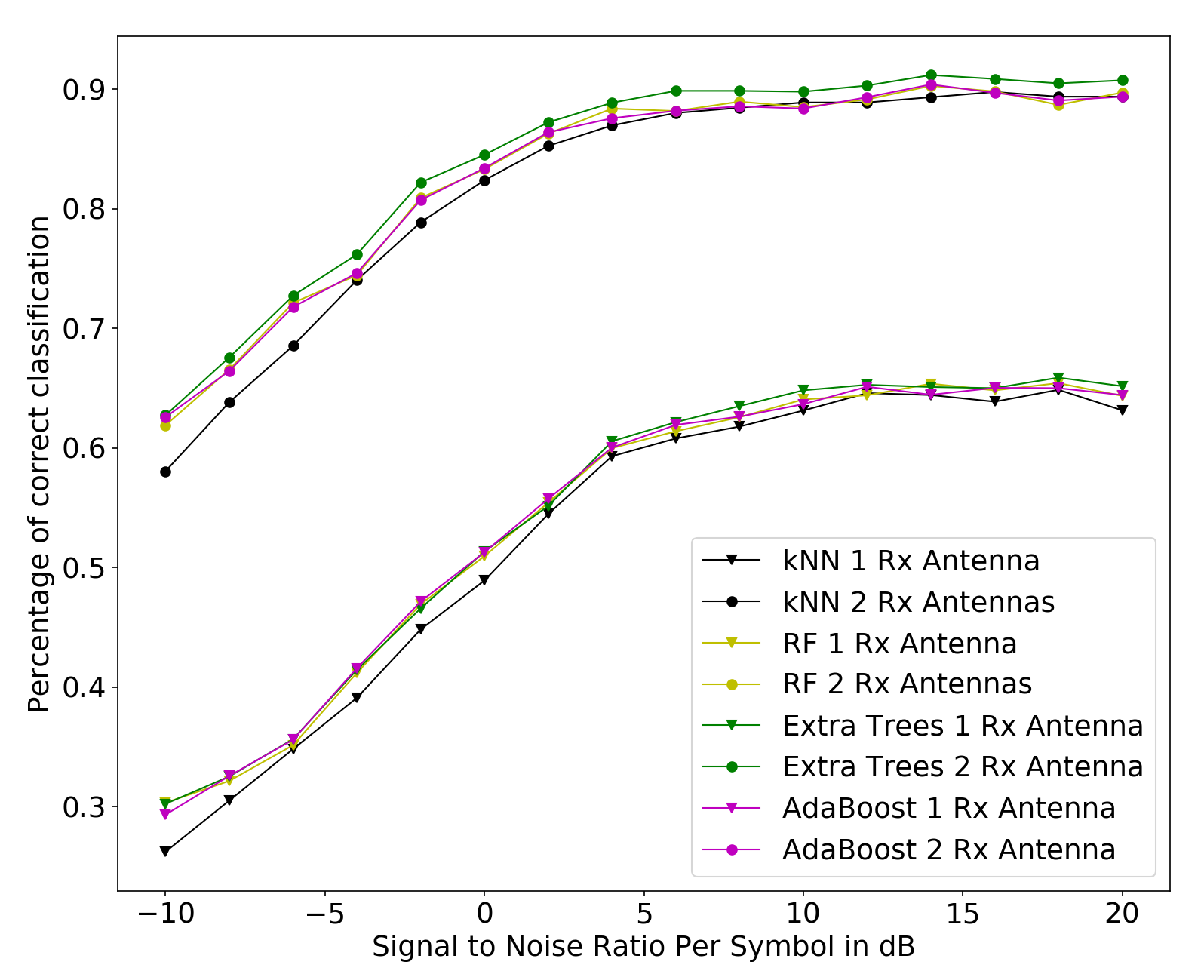}
\caption{Percentage of correct classification of HMC/UAC.}
\label{fig:plot_universal}
\end{figure}

\begin{table}[t]
\centering
\begin{tabular}{|l|l|l|l|l|l|} 
 \hline
 Dataset&Classifier & -10dB & 0dB & 10dB & 20dB \\ [0.5ex] 
 \hline
 \multirow{4}{*}{$N_R$=1} 
 &kNN & 26.22  & 48.91  & 63.12  &  63.14  \\
 &RF & 30.32  & 50.94  & 64.05  & 64.35  \\
 &ET & 30.23  & 51.34  & 64.81  & 65.16  \\
 &AdaBoost & 29.32  & 51.27  & 63.57  & 64.42 \\
  \hline

 \multirow{4}{*}{$N_R$=2} 
 & kNN & 58.05  & 82.38  & 88.88  &  89.37  \\
 & RF & 61.89  & 83.31  & 88.49  & 89.72  \\
 & ET & 62.73  & 84.51  & 89.79  & 90.74  \\
 & AdaBoost & 62.56  & 83.4  & 88.37  & 89.37 \\
 \hline
 \end{tabular}
 \caption{Classification accuracy of the HMC/UAC combination.}
\label{table:acc_jont_universal_1ant}
\end{table}

\begin{table}[t]
	\centering
	\begin{tabular}{|l|l|l|l|l|l|} 
		\hline
		Dataset&Classifier & -10dB & 0dB & 10dB & 20dB \\ [0.5ex] 
		\hline
		\multirow{4}{*}{$N_R$=1} 
		&kNN & 27.29  & 50.87  & 64.18  &  64.44  \\
		&RF & 31.08  & 54.6  & 66.18  & 66.22  \\
		&ET & 31.08  & 54.32  & 66.22  & 66.82  \\
		&AdaBoost & 31.27  & 54.6  & 66.18  & 66.87 \\
		\hline
		
		\multirow{4}{*}{$N_R$=2} 
		&kNN & 59.05  & 82.98  & 89.23  &  89.72  \\
		&RF & 64.42  & 85.39  & 90.55  & 91.01  \\
		&ET & 64.32  & 86.22  & 91.04  & 91.73  \\
		&AdaBoost & 64.21  & 85.6  & 90.46  & 90.94 \\
		\hline
	\end{tabular}
	\caption{Classification accuracy of the HMC/DeAC combination.}
	\label{table:acc_jont_dedicated_2ant}
\end{table}

\begin{figure}[h!]
\centering
 \includegraphics[width = \linewidth,]{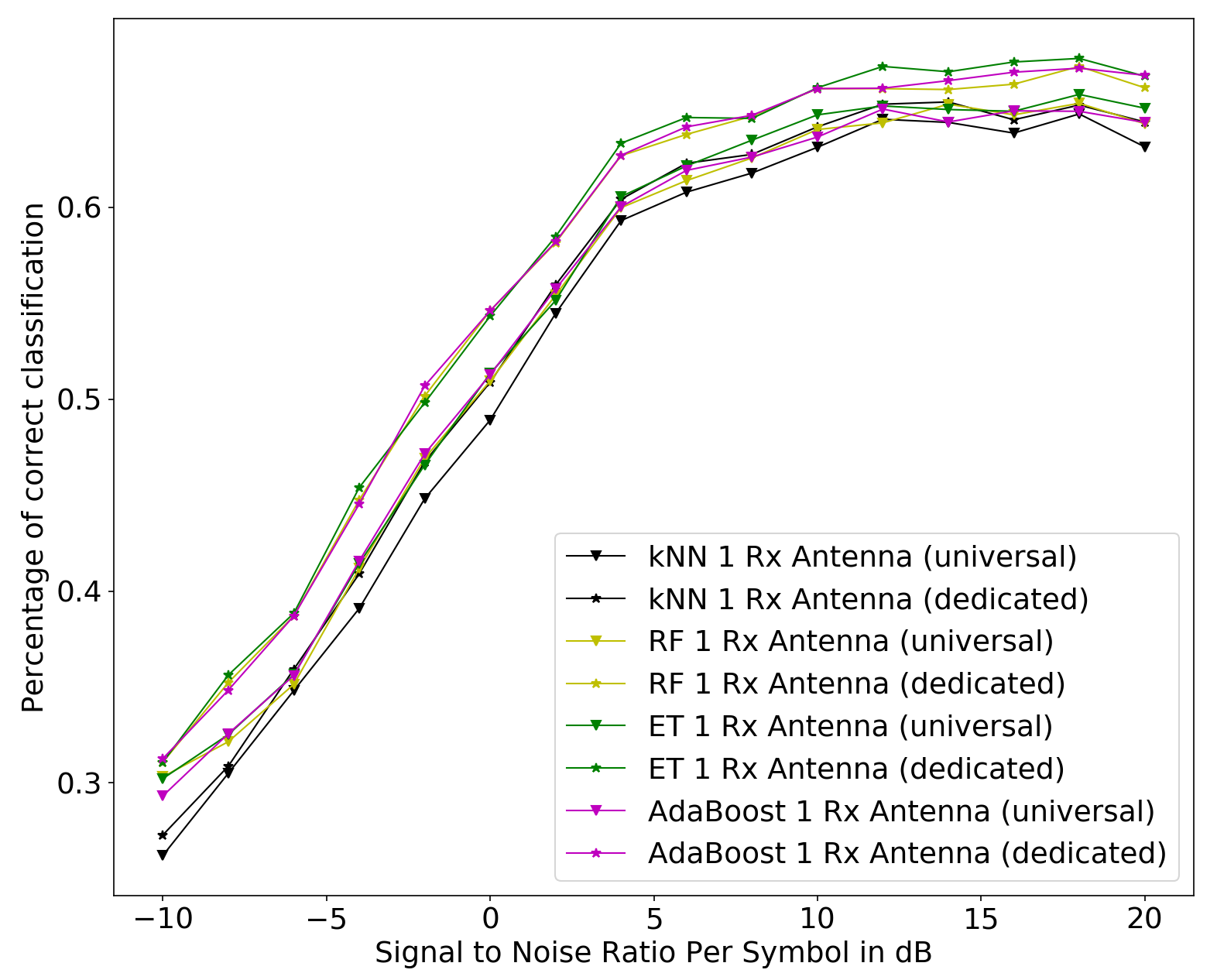}
\caption{Results for joint $N_T$ and $\mathcal{M}$ classification. Comparison of the Universal and Dedicated classifiers for $N_R$=1.}
\label{fig:plot_dedicated_vs_universal1}
\end{figure}

\begin{figure}[h!]
\centering
 \includegraphics[width =\linewidth]{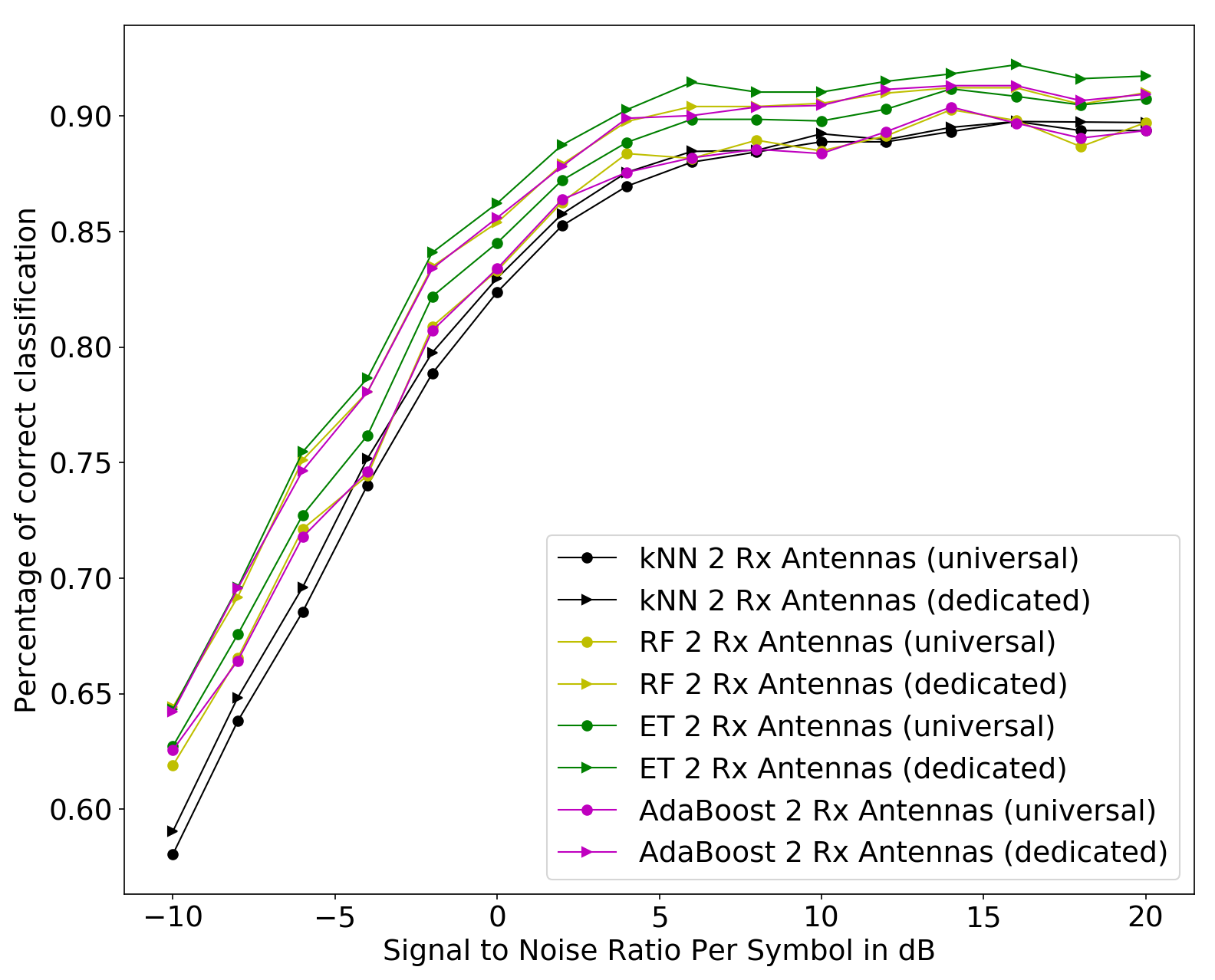}
\caption{Results for joint $N_T$ and $\mathcal{M}$ classification. Comparison of the Universal and Dedicated classifiers for $N_R$=2.}
\label{fig:plot_dedicated_vs_universal2}
\end{figure}

Starting with the scenario involving the \textit{Universal} classifier, the contribution of multiple antennas is evident as can be seen in Fig. \ref{fig:plot_universal}. From the results presented in the previous sections, it was expected that the use of two antennas at the receiver would result in a rise of the classification accuracy in the joint scenario as well. For once more, the use of two Rx antennas improved the performance by a percentage up to 26\%, with the most notable improvement spotted in low SNR values. For example, the kNN algorithm achieves an accuracy score of 26.22\% in $-10$dB and with $N_T$=1, while the accuracy reaches up to 58.5\% in the same SNR regime with $N_R$=2. Experiments do not show any particular advantage in using a certain classification algorithm over the others, since their performance is almost identical. The only exception is kNN, which seems to under-perform slightly for lower SNR. The reader may refer to Table \ref{table:acc_jont_universal_1ant} for a more detailed report on the classification accuracy for $N_R$=1 and $N_R$=2 respectively. In the second scenario, where dedicated classifiers for each modulation are present, the same behavior can be observed regarding the contribution of multiple Rx antennas (that is higher $N_R$ antennas leads to better classification accuracy). 
It is interesting to examine is the improvement that the use of the dedicated classifiers induces to the overall classification performance. In Fig. \ref{fig:plot_dedicated_vs_universal1} and \ref{fig:plot_dedicated_vs_universal2} it can be seen that indeed by using dedicated classifiers, a slightly higher classification accuracy can be achieved  Around 2,5-3\% regardless of the SNR regime.  

\begin{figure}[t]
\centering
 \includegraphics[width = \linewidth,]{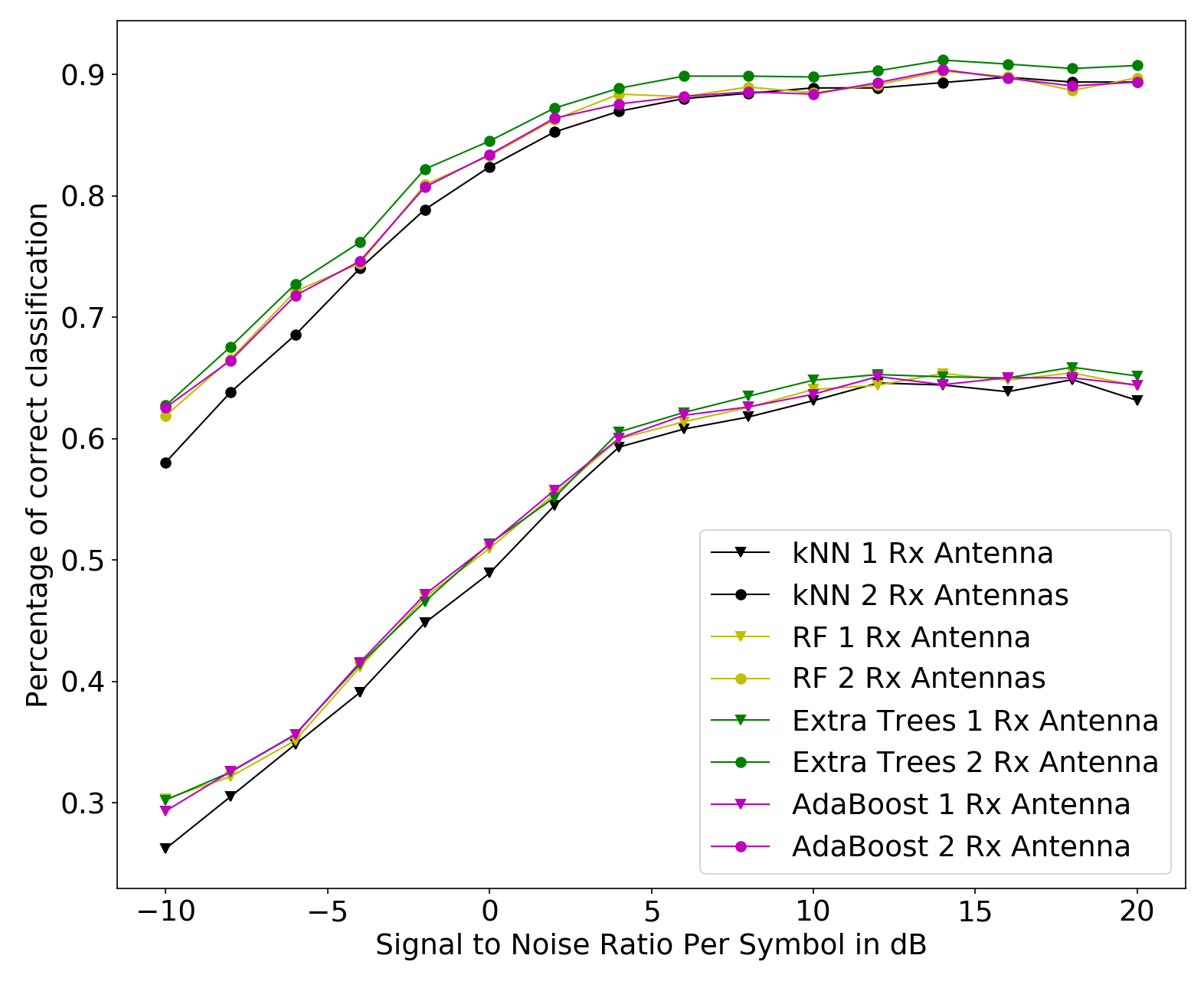}
\caption{Percentage of correct classification for the HMC/DeAC combination.}
\label{fig:plot_dedicated}
\end{figure}

\emph{In this section we showed that we can achieve very high accuracy for the problem of joint modulation and antenna classification. When we treated the two problems sequentially, that is when we used a Modulation classifier first and matched its results to a Dedicated Antenna Classifier (with pre-trained models to each specific modulation set $\mathcal{M}$), namely the HMC/DeAC combination, we observed slightly superior performance to the scheme where the problem is dealt joint and in parallel (HMC/UAC).}
\section{Conclusions}
\label{sec:conclusions}
In this paper, we studied the problem of modulation and antenna classification in digital wireless transmission. The two classification sub-problems were initially examined separately. It was shown that the classification accuracy is increased with SNR, number of Rx antennas, and the use of ensemble classification methods. For the joint classification problem we followed two approaches. One, where the sub-problems are solved independently and in parallel, and one where the antenna classifier waits on the result of the modulation classifier. The later scheme performed slightly better under all case of SNR, number of Rx antennas, and classification algorithms. The proposed schemes do not require any knowledge/details of the used modulation schemes and the way the Tx antennas are used as it is fully data-driven and not model-driven. The results of our approach pave the way for more ML-based data-driven techniques that reveal more characteristics of the Tx.
\bibliographystyle{IEEEtran}
\bibliography{references}

\begin{thebibliography}{10}
\providecommand{\url}[1]{#1}
\csname url@samestyle\endcsname
\providecommand{\newblock}{\relax}
\providecommand{\bibinfo}[2]{#2}
\providecommand{\BIBentrySTDinterwordspacing}{\spaceskip=0pt\relax}
\providecommand{\BIBentryALTinterwordstretchfactor}{4}
\providecommand{\BIBentryALTinterwordspacing}{\spaceskip=\fontdimen2\font plus
\BIBentryALTinterwordstretchfactor\fontdimen3\font minus
  \fontdimen4\font\relax}
\providecommand{\BIBforeignlanguage}[2]{{%
\expandafter\ifx\csname l@#1\endcsname\relax
\typeout{** WARNING: IEEEtran.bst: No hyphenation pattern has been}%
\typeout{** loaded for the language `#1'. Using the pattern for}%
\typeout{** the default language instead.}%
\else
\language=\csname l@#1\endcsname
\fi
#2}}
\providecommand{\BIBdecl}{\relax}
\BIBdecl

\bibitem{hameed2009likelihood}
F.~Hameed, O.~A. Dobre, and D.~C. Popescu, ``On the likelihood-based approach
  to modulation classification,'' \emph{IEEE Transactions on Wireless
  Communications}, vol.~8, no.~12, pp. 5884--5892, 2009.

\bibitem{dobre2007survey}
O.~A. Dobre, A.~Abdi, Y.~Bar-Ness, and W.~Su, ``Survey of automatic modulation
  classification techniques: classical approaches and new trends,'' \emph{IET
  communications}, vol.~1, no.~2, pp. 137--156, 2007.

\bibitem{assaleh1992new}
K.~Assaleh, K.~Farrell, and R.~Mammone, ``A new method of modulation
  classification for digitally modulated signals,'' in \emph{MILCOM 92
  Conference Record}.\hskip 1em plus 0.5em minus 0.4em\relax IEEE, 1992, pp.
  712--716.

\bibitem{dan2005new}
{Wu Dan}, {Gu Xuemai}, and {Guo Qing}, ``A new scheme of automatic modulation
  classification using wavelet and wsvm,'' in \emph{2nd Asia Pacific Conference
  on Mobile Technology, Applications and Systems}, 2005.

\bibitem{xie2008efficient}
F.~Xie, C.~Li, and G.~Wan, ``An efficient and simple method of mpsk modulation
  classification,'' in \emph{4th International Conference on Wireless
  Communications, Networking and Mobile Computing}.\hskip 1em plus 0.5em minus
  0.4em\relax IEEE, 2008, pp. 1--3.

\bibitem{peng2018modulation}
S.~Peng, H.~Jiang, H.~Wang, H.~Alwageed, Y.~Zhou, M.~M. Sebdani, and Y.-D. Yao,
  ``Modulation classification based on signal constellation diagrams and deep
  learning,'' \emph{IEEE transactions on neural networks and learning systems},
  vol.~30, no.~3, pp. 718--727, 2018.

\bibitem{ahmadi2010using}
N.~Ahmadi, ``Using fuzzy clustering and ttsas algorithm for modulation
  classification based on constellation diagram,'' \emph{Engineering
  Applications of Artificial Intelligence}, vol.~23, no.~3, pp. 357--370, 2010.

\bibitem{zhou2010signal}
X.~Zhou, Y.~Wu, and B.~Yang, ``Signal classification method based on support
  vector machine and high-order cumulants.'' \emph{Wirel. Sens. Netw.}, vol.~2,
  no.~1, pp. 48--52, 2010.

\bibitem{swami2000hierarchical}
A.~Swami and B.~M. Sadler, ``Hierarchical digital modulation classification
  using cumulants,'' \emph{IEEE Transactions on communications}, vol.~48,
  no.~3, pp. 416--429, 2000.

\bibitem{liu2006novel}
L.~Liu and J.~Xu, ``A novel modulation classification method based on high
  order cumulants,'' in \emph{International Conference on Wireless
  Communications, Networking and Mobile Computing}.\hskip 1em plus 0.5em minus
  0.4em\relax IEEE, 2006, pp. 1--5.

\bibitem{su2013feature}
W.~Su, ``Feature space analysis of modulation classification using very
  high-order statistics,'' \emph{IEEE Communications Letters}, vol.~17, no.~9,
  pp. 1688--1691, 2013.

\bibitem{dobre2003higher}
O.~A. Dobre, Y.~Bar-Ness, and W.~Su, ``Higher-order cyclic cumulants for high
  order modulation classification,'' in \emph{IEEE Military Communications
  Conference, 2003. MILCOM 2003.}, vol.~1.\hskip 1em plus 0.5em minus
  0.4em\relax IEEE, 2003, pp. 112--117.

\bibitem{o2017introduction}
T.~O’Shea and J.~Hoydis, ``An introduction to deep learning for the physical
  layer,'' \emph{IEEE Transactions on Cognitive Communications and Networking},
  vol.~3, no.~4, pp. 563--575, 2017.

\bibitem{meng2018automatic}
F.~Meng, P.~Chen, L.~Wu, and X.~Wang, ``Automatic modulation classification: A
  deep learning enabled approach,'' \emph{IEEE Transactions on Vehicular
  Technology}, vol.~67, no.~11, pp. 10\,760--10\,772, 2018.

\bibitem{wang2020lightamc}
Y.~Wang, J.~Yang, M.~Liu, and G.~Gui, ``Lightamc: Lightweight automatic
  modulation classification via deep learning and compressive sensing,''
  \emph{IEEE Transactions on Vehicular Technology}, vol.~69, no.~3, pp.
  3491--3495, 2020.

\bibitem{zhang2018automatic}
D.~Zhang, W.~Ding, B.~Zhang, C.~Xie, H.~Li, C.~Liu, and J.~Han, ``Automatic
  modulation classification based on deep learning for unmanned aerial
  vehicles,'' \emph{Sensors}, vol.~18, no.~3, p. 924, 2018.

\bibitem{ramjee2019fast}
S.~Ramjee, S.~Ju, D.~Yang, X.~Liu, A.~E. Gamal, and Y.~C. Eldar, ``Fast deep
  learning for automatic modulation classification,'' \emph{arXiv preprint
  arXiv:1901.05850}, 2019.

\bibitem{ali2017unsupervised}
A.~Ali and F.~Yangyu, ``Unsupervised feature learning and automatic modulation
  classification using deep learning model,'' \emph{Physical Communication},
  vol.~25, pp. 75--84, 2017.

\bibitem{o2016convolutional}
T.~J. O’Shea, J.~Corgan, and T.~C. Clancy, ``Convolutional radio modulation
  recognition networks,'' in \emph{International conference on engineering
  applications of neural networks}.\hskip 1em plus 0.5em minus 0.4em\relax
  Springer, 2016, pp. 213--226.

\bibitem{abdelmutalab2016automatic}
A.~Abdelmutalab, K.~Assaleh, and M.~El-Tarhuni, ``Automatic modulation
  classification based on high order cumulants and hierarchical polynomial
  classifiers,'' \emph{Physical Communication}, vol.~21, pp. 10--18, 2016.

\bibitem{oularbi2012enumeration}
M.-R. Oularbi, S.~Gazor, A.~Aissa-El-Bey, and S.~Houcke, ``Enumeration of base
  station antennas in a cognitive receiver by exploiting pilot patterns,''
  \emph{IEEE communications letters}, vol.~17, no.~1, pp. 8--11, 2012.

\bibitem{oularbi2013exploiting}
------, ``Exploiting the pilot pattern orthogonality of ofdma signals for the
  estimation of base stations number of antennas,'' in \emph{2013 8th
  International Workshop on Systems, Signal Processing and their Applications
  (WoSSPA)}.\hskip 1em plus 0.5em minus 0.4em\relax IEEE, 2013, pp. 465--470.

\bibitem{ohlmer2008algorithm}
E.~Ohlmer, T.-J. Liang, and G.~Fettweis, ``Algorithm for detecting the number
  of transmit antennas in mimo-ofdm systems: receiver integration,'' in
  \emph{2008 IEEE 68th Vehicular Technology Conference}.\hskip 1em plus 0.5em
  minus 0.4em\relax IEEE, 2008, pp. 1--5.

\bibitem{somekh2007detecting}
O.~Somekh, O.~Simeone, Y.~Bar-Ness, and W.~Su, ``Detecting the number of
  transmit antennas with unauthorized or cognitive receivers in mimo systems,''
  in \emph{MILCOM 2007-IEEE Military Communications Conference}.\hskip 1em plus
  0.5em minus 0.4em\relax IEEE, 2007, pp. 1--5.

\bibitem{turan2015joint}
M.~Turan, M.~{\"O}ner, and H.~A. {\c{C}}{\i}rpan, ``Joint modulation
  classification and antenna number detection for mimo systems,'' \emph{IEEE
  Communications Letters}, vol.~20, no.~1, pp. 193--196, 2015.

\bibitem{tse2005fundamentals}
D.~Tse and P.~Viswanath, \emph{Fundamentals of wireless communication}.\hskip
  1em plus 0.5em minus 0.4em\relax Cambridge university press, 2005.

\bibitem{argyriou2020}
\BIBentryALTinterwordspacing
A.~Argyriou, ``Secrecy in wireless communication through closed-loop receiver
  de-synchronization,'' \emph{Physical Communication}, vol.~43, p. 101195,
  2020. [Online]. Available:
  \url{http://www.sciencedirect.com/science/article/pii/S187449072030272X}
\BIBentrySTDinterwordspacing

\bibitem{geisinger2010classification}
N.~P. Geisinger, ``Classification of digital modulation schemes using linear
  and nonlinear classifiers,'' NAVAL POSTGRADUATE SCHOOL MONTEREY CA, Tech.
  Rep., 2010.

\bibitem{cover1967nearest}
T.~Cover and P.~Hart, ``Nearest neighbor pattern classification,'' \emph{IEEE
  transactions on information theory}, vol.~13, no.~1, pp. 21--27, 1967.

\bibitem{geurts2006extremely}
P.~Geurts, D.~Ernst, and L.~Wehenkel, ``Extremely randomized trees,''
  \emph{Machine learning}, vol.~63, no.~1, pp. 3--42, 2006.

\end{thebibliography}

\end{document}